\documentclass[11pt]{JHEP3} 

\JHEPspecialurl{http://jhep.sissa.it/JOURNAL/JHEP3.tar.gz}

\usepackage{epsfig,multicol,bbm}
\usepackage{amsmath}
\usepackage{amsfonts}
\usepackage{amssymb}
\usepackage{eso-pic}
\newcommand\fverb{\setbox\fverbbox=\hbox\bgroup\verb}
\newcommand\fverbdo{\egroup\medskip\noindent%
			\fbox{\unhbox\fverbbox}\ }
\newcommand\fverbit{\egroup\item[\fbox{\unhbox\fverbbox}]}
\newbox\fverbbox

\def\beq{\begin{eqnarray}}
\def\eeq{\end{eqnarray}}
\def\be{\begin{equation}}
\def\ee{\end{equation}}

\def\={\triangleq}
\def\nn{\nonumber\\}

\def\ub{\underleftarrow}

\def\mbf{\mbox{\boldmath}}

\def\lie{\pounds}
\def\l{\ell}

\def\bm{\bar m}

\def\bA{{\bf A}}
\def\bF{{\bf F}}

\def\N{\mathcal{N}}

\title{Entropy of Black Holes in $\mathcal{N}=2$ Supergravity}

\author{Ayan Chatterjee,\\ Theory Division, Saha Institute of
Nuclear Physics,\\ 1/AF, Bidhannagar, Kolkata 700064, India.\\
	E-mail: \email{ayan.chatterjee@saha.ac.in}}

 \preprint{Aaaa/Mm/Yy}

\abstract{Using the techniques of the isolated horizon formalism, we construct 
space of solutions of asymptotically flat extremal black holes
in $\mathcal{N}=2$ pure supergravity in $4$ dimensions. We prove the
laws of black hole mechanics. Further, restricting to constant area phase space,
we show that the spherical horizons admit a $U(1)$ Chern- Simons theory.
Standard way of quantizing this topological theory and counting states 
confirms that entropy is indeed proportional to the area of horizon.}


\begin{document} 

\section{Introduction}
Black holes are ``the simplest macroscopic objects''
made out of spacetime \cite{Chandra}. This stimulates the hope that black holes
will turn out to provide crucial clues for a quantum theory of gravity just
as the hydrogen atom helped us unravel the secrets of the atomic
system. The most important development in the last few decades was to establish
that black holes behave as macroscopic states in thermal equilibrium and their
dynamical laws (laws of black hole mechanics) being similar to laws of
thermodynamics \cite{bch}. This observation prompted Bekenstein and Hawking
to argue that black holes
indeed have temperature related to the surface gravity and their entropy 
is related to the area of the black hole horizon \cite{b,h}. However, one
needs statistical interpretation for such \emph{thermodynamic}
arguments. It is expected that any quantum theory of gravity should be able to
specify the microstates of the black hole spacetime and the leading term in
Boltzmann definition of entropy would be proportional to the area of the
horizon. Supergravity and string theories are leading candidates of quantum
theory. Black holes occurring in these theories are subjects of intense
study. Moreover, these solutions can be interpreted
as self-gravitating solitons interpolating between different vacua of the theory
\cite{GT,S_97,DK}. The extremal Reissner-Nordstrom solution arising in pure
$\mathcal{N}=2$ supergravity in $4$-dimensions is the simplest example
\cite{GH1,BMG,Tod1,Tod2, FKS}. The solitons in string theory (and low energy
effective actions) play a central
role in understanding string dualities. Alternatively, string theory has been
used to investigate quantum properties of extremal black holes
\cite{stro,mal,peet,dha,sen}. Extremal black holes are BPS solitons as they have
some residual supersymmetry of the extended supersymmetric theories (see
\cite{MSW} for other interpretations of the term BPS). Using this
BPS property, one can do computations like entropy for example, in perturbative
regime and extrapolate to non-perturbative region. These BPS solutions have
high degree of supersymmetry as isometries and this shields the counting of
states over large variation of modular parameters. More solutions of the BPS
type exist in higher dimensions with various degree of supersymmetry. They have 
proved to be of great interest for establishing various subtleties related to
the entropy calculation (see \cite{m,o,y,fre,ste}).

A popular way of finding the `macroscopic' entropy of black holes (in different
theories of gravity) is to use the Killing horizon formalism and the Wald
formula \cite{w,wi}. However, it is well known that this approach is difficult
to implement convincingly for extremal black holes (see \cite{j,
cg_holst}). Indeed, the laws of mechanics depend on the assumption that the
Killing horizon admit \emph{bifurcation spheres} (see \cite{w,rw}). These are
special spheres which lie in the intersection of the past and the future
horizon. Wald's proposal for interpreting the Noether charge (corresponding to
the diffeomorphism invariance associated with the Killing vector field
generating the horizon) as the ``entropy" crucially depends on the existence of
the bifurcation sphere \cite{w,wi,j}. But, extremal black holes are not past
complete and hence do not have any bifurcation sphere. This implies that 
the proof of the laws of mechanics and the determination of the entropy of these
black holes via the Wald formula remains a suspect. One might consider 
taking extremal limit of the entropy of corresponding non-extremal black holes.
Arguments presented above also question such procedure. Consider the space of
solution (of any theory of gravity) containing Killing horizons. If the laws of
black hole mechanics hold in this phase space, then it does not contain extremal
black holes. The extremal black holes do not exist as a limit 
point in this phase space. In other words, no sequence of solution can
be constructed in this space of solution which will have their surface 
gravity limiting to zero. So in this phase space, taking \emph{extremal limit}
of the Noether charge for non-extremal black holes remains ill defined.
There are at least two ways to solve this problem:
first, to find a way to deal with extremal black holes in the Wald formulation
or second, to consider the isolated horizon formalism  which supports extremal
horizons in its phase space. Our aim in this paper is to show that the second
possibility can be naturally used to calculate the entropy of spherical black
holes in pure $\mathcal{N}=2$ supergravity. 

Isolated horizon (IH) \cite{abf_let, abf, a_prl, abl_rot, abl_generic, ak_lrr,
cg1} is a local definition of black hole horizon. Unlike event horizons or the
Killing horizon, this definition does not require the knowledge of spacetime
external to the horizon. The knowledge of the entire spacetime is redundant . IH
boundary conditions are the minimal set of conditions that any generic black
hole horizon (extremal/nonextremal) is expected to satisfy. The most important
characterization of IH is that they are expansion-free. This condition 
separates an arbitrary null surface from a black hole horizon. For example,
the Minkowski light cone expands in the future (and also in the past) and hence
is not an IH. This is expected because the Minkowski null-cone behaves as a
horizon only for the Rindler observers. The condition of expansion-freeness
implies that no matter field falls inside the horizon. Moreover, the boundary
conditions imply that there exists Killing vectors fields \emph{on the black
hole horizon only}. Thus there might be radiation arbitrarily close to the
horizon but are not allowed to cross it. This allows a large class of
solutions to satisfy the IH boundary conditions. Indeed, the space of solutions
of any theory of gravity admitting an IH as an inner  boundary is larger than
that of the Killing or the event horizon (these definitions require
Killing vector fields outside the horizon too). The original formulation of IH
however, had extremal and non- extremal horizons in distinct phase spaces. The
Weak Isolated Horizon (WIH) uses a set of milder conditions than IH
and puts these two classes of solutions in equal footing \cite{cg1,
cg_holst}. The IH (or the WIH) formulation does not require bifurcation spheres
to establish the laws of black holes mechanics or to determine the entropy. In
other words, the IH (or the WIH) formlation provides the ideal set-up to study
mechanics of extremal black hole horizons arising in string theory or
supergravity. The IH formulation is also useful to compute the entropy of
horizons in loop quantum gravity (LQG) approach. The point of view 
of this approach is that the horizon supports the effective degrees of
freedom that arise out of a well defined interaction between the bulk and
the boundary configurations. These microstates residing on the boundary
capture all the essential features of the spacetime. In other words, the
microstates relevant for entropy counting are localized on the horizon only.
It is also not very difficult to guess the nature of the theory on
horizon. Since the surface is null, it does not support a metric theory. 
It is only natural that the effective theory on this null surface be a
topological theory. The theory turns out to be a $U(1)$ Chern-Simons theory
\cite{ack}. Entropy of the horizon can then be obtained by directly
geometrically quantizing this Chern-Simons theory \cite{abck, abk}.
Alternatively, the authors of \cite{km,kmd} used techniques of conformal field
theory to obtain the entropy and corrections to all orders in Planck length.

We shall use the IH formalism approach to deal with the black hole
solutions in $\mathcal{N}=2$ supergravity. In \cite{cg1}, it was proposed that
the IH formalism introduces an  ideal set-up to study black hole solutions in
string theory and supergravity. In \cite{bl1,bl2,l2}, the authors formulated the
precise notions required to make IH amiable to supergravity solutions. However,
the detail compatibility study between the IH boundary conditions and
the conditions arising from the supergravity theory is still to be carried out.
Moreover, the derivation of the laws of black hole mechanics and the
calculation of entropy (as is done for IH in GR) are needed for detailed
understanding of the non-perturbative aspects of spacetime. The aim of the
present paper is to fill this gap.

We will be interested in classical spacetimes which are 
purely bosonic solutions of supergravity theories. The solutions which retain
some supersymmetry of the theory are lebelled BPS saturated. The BPS condition
bounds the mass (measured at infinity) from below by a function of the
asymptotic charges of the fields (for $\mathcal{N}=2$, by charges of the Maxwell
field) (however, see \cite{MSW}). When the bound is attained, the
classical spacetime admits \emph{Killing spinor fields}. The supersymmetry
transformations generated by these Killing spinors are such that the bosonic
fields are left invariant while the supersymmetry transformations of
the fermionic fields vanish. These conditions on the fields can be  
turned into a set of first order differential equation called the Killing
spinor equations (KSE)\footnote{The KSEs are different for different theories
but since we are interested in classical configuration, it remains true for
any other set of fermion field}. Alternatively, given the KSEs for a
supergravity theory, their solution leads to classical configurations
with unbroken supersymmetry. However, only a few of these configurations
actually solve the corresponding supergravity equations of motion. Solutions of
KSEs for different supergravity theories are of great interest.

Our main interest in this paper is to study the black holes in $\mathcal{N}=2$
supergravity using the IH formalism. As a concrete example,
we shall consider the extremal Reissner-Nordstrom black hole solution
in $\mathcal{N}=2$ supergravity. This is also a well solution in the 
Einstein-Maxwell theory which in fact, is a consistent truncation of the
$\mathcal{N}=2$  supergravity. For these global solutions, the mass (and charge)
measured by an asymptotic observer equals the mass (and charge) defined on the
horizon. Thus, this particular solution is supersymmetric or BPS for any
observer, at asymptopia or at the horizon. These kind of solutions will be
called globally supersymmetric. In the context of the IH formalism,
the most general cases are those where one has access \textit{only} to mass and
charge defined locally on the horizon and does not have any knowledge of the
nature of the exterior spacetime. In that case, it is natural to consider
solutions (or configurations) which saturate the BPS bound solely on the
horizon, \emph{i.e.} local horizon mass equal the charge of the field equated on
the horizon. Indeed, IH formalism can incorporate solutions which are
supersymmetric on the horizon \textit{only} while the bulk spacetime may have no
residual supersymmetry of the theory. We repeat that, while it is enough for the
IH formalism to require KSEs to hold just on the horizon, the configurations
which solve the KSEs globally (like the extremal RN solution) will also
naturally be part of the IH phase space.

As mentioned before, we shall  investigate the applicability
of the IH formalism for black hole solutions arising in supergravity. 
We will use the global and supersymmetric Reissner-Nordstrom solution arising
in pure $\mathcal{N}=2$ supergravity for consistency study. First, we need to
check that in the region outside the horizon (when the Killing vector is
timelike), the solution of KSEs (equations arising because of BPS condition)
imply a Reissner-Nordstrom like configuration \emph{i.e.} static with
asymptotically flat geometry, invariant under half of the the $\mathcal{N}=2$
supersymmetries. Second, when the Killing vector is null, for e.g. on the
horizon, the KSEs give rise to configurations whose geometric structures are
consistent with the ones derived from IH boundary conditions.\footnote{This kind
of construction can be also done for general relativity. For example, we
might want to construct configurations which solve the Killing equation
 $\nabla_{(a}t_{b)}=0$ where $t^a$ is a timelike vector field. The static
configurations includes the extremal Reissner-Nordstrom like spacetime. However,
not all of these configurations are solutions of the Einstein equation. The
solutions are those for which the constants $M$ and $Q$ in the configuration
can be identified with mass and charge respectively (for $M=Q$, we get the
extremal Reissner-Nordstrom solution). In case $t^a$ becomes
null (for example on the horizon), the solutions of the Killing equation will
comprise of configurations which satisfy the IH boundary conditions.} We shall
call a horizon \emph{supersymmetric weak isolated horizon} (SWIH) if the
conditions for existence of Killing spinors on the horizon are compatible with
the IH boundary conditions. In other words, on a SWIH, the KSEs arising
because of the BPS nature of the horizon will be consistent with the IH
boundary conditions. Once the SWIH is defined, the next task is to
construct the phase-space of the theory of gravity in hand  with appropriate
boundary conditions. Our phase space will consist of all solutions of 
$\mathcal{N}=2$ supergravity which satisfy the SWIH boundary condition at
the horizon and are asymptotically flat at infinity.\footnote{It is important to
note that the phase space will only consist of the solutions of the equations
of motion of $\mathcal{N}=2$ supergravity. These are a subset of all 
solutions of the KSEs of the given supergravity theory.} In this paper, we shall
use the generalization of the Holst action \cite{hol} as the action for
$\mathcal{N}=2$ pure supergravity \cite{k}. Using well known techniques of
covariant phase space, one can construct the symplectic structure on this SWIH
phase space \cite{abr,lw}. The first law for the SWIH will then follow from this
symplectic structure. It will also follow from this symplectic
structure that the topological theory on fixed area phase space of SWIH is
a $U(1)$ Chern-Simons theory.

The plan of this paper is as follows: First, we spell out the isolated boundary
conditions to be imposed on a generic null surface. Second, we study the
constraints arising from Killing spinor equations and show that they are
consistent with IH boundary conditions. Thirdly, we construct the
space of solution of $\mathcal{N}=2$ pure supergravity which satisfy the SWIH
boundary conditions at horizon and are asymptotically flat at infinity. 
We construct the symplectic structure and prove the laws of black hole
mechanics. Next, we shall go to fixed area phase space and 
and identify the $U(1)$ Chern-Simons theory as the boundary theory.

\section{Isolated horizon boundary conditions}
We consider a $4$- dimensional spacetime manifold $\mathcal{M}$
equipped with a metric $g_{ab}$ having Lorentzian signature
$(-, +,+, +)$ and a null hypersurface $\Delta$.
Let $\ell^a$ be a future directed null normal on $\Delta$. However, if $\ell^a$
is a future directed null normal, then so is $\xi\ell^a$, where $\xi$ is any
positive function on $\Delta$. Two null normals $\ell^a$ and $\ell^{'a}$ on
$\Delta$ will be called equivalent if $\ell^{'a}=\xi\ell^a$. Thus, $\Delta$
naturally admits an equivalence class of null normals.  We shall
denote this equivalence class by $[\,\xi\ell^{a}\,]$. Let
us denote by $q_{ab}\triangleq g_{\ub{ab}}$ the degenerate intrinsic metric on
$\Delta$ which is induced by the spacetime metric $g_{ab}$ (indices that are not
intrinsic on $\Delta$ will be pulled back, denoted by an arrow under them, and
$\triangleq$ signifies that the equality holds {\em only on} $\Delta$). Thus
$q_{ab}$ has a signature $(0, +, +)$. A tensor $q^{ab}$ will be called an {\em
inverse} of $q_{ab}$ if it obeys the condition $q^{ab}q_{ac}q_{bd}\triangleq
q_{cd}$. The inverse metric $q^{ab}$, however, is not unique as one can redefine
it as $q^{ab}\mapsto q^{ab}+k^{(a}\ell^{b)}$, where $k^{a}$ is any vector field
tangential with $\Delta$. The expansion $\theta_{(\ell\,)}$ of the null normal
$\ell^{a}$ is then defined by $\theta_{(\ell\,)}=q^{ab}\nabla_a\ell_b$, where
$\nabla_a$ is the spacetime covariant derivative compatible with $g_{ab}$. Note
that the expansion $\theta_{(\ell\,)}$ is insensitive to the ambiguity in the
inverse metric but it varies under the scaling of the null normal
$\ell^a$ in the equivalence class $[\xi\ell^a]$ by
$\theta_{(\xi\ell\,)}=\xi\theta_{(\ell\,)}$.

In what follows, we shall work with the Newmann-Penrose (NP) null tetrad basis
$(\ell^{a},n^{a}, m^{a}, {\bar m}^{a})$, $n_{a}$ being normal to the foliation
of $\Delta$ by $S^2$ and $m^a, \bar m^{a}$ are tangential to $2$- spheres. The
basis vector obey the orthonormality conditions $\l.n=-1=-m.\bar m$, others 
being zero. This is specially suited for the present study because one of the
null-normals in the equivalence class $[\xi\ell^a]$ coencide with the basis
vector $\ell^a$. Moreover, in this basis, many components of the connection
vanish making the calculations much simpler than that in the coordinate basis.
For future calculations with supergravity, it will be convenient to use the
spinor basis alongside \cite{PR_book,abf, ack}. We can find a spin dyad
$(\iota^{A}, o^{A})$ with normalization condition $\iota^{A}o_{A}\=1$. The null
vectors $(l,n, m, \bar m)$ are related to these dyad basis by the following
relations:
\begin{eqnarray}\label{lnmdef}
\l^{a}&\=&i\sigma^{a}{}_{AA^{'}}o^{A}\bar{o}^{A^{'}}~~n_{a}\=i\sigma_{a}{}^{AA^{
'}}\iota_{A}\bar{\iota}_{A^{'}}\nonumber\\
m^{a}&\=&i\sigma^{a}{}_{AA^{'}}o^{A}\bar{\iota}^{A^{'}}~~\bar
m^{a}\=i\sigma^{a}{}_{AA^{'}}\bar{o}^{A^{'}}\iota^{A},
\end{eqnarray}
where $\sigma^{a}{}_{AA^{'}}$ is called the soldering form.

The isolated horizon boundary conditions for spherically symmetric
cases can be stated in terms of the spin dyads as follows \cite{abf, ack}. The
surface $\Delta$ will be called a non-expanding horizon (NEH) if 
\begin{enumerate}\label{ihbc}
\item $\Delta$ is topologically $S^{2}\times R$.
\item The spin dyads $(o_{a}, \iota_{A})$ are constrained to satisfy
\footnote{Quantities which are not intrinsic to $\Delta$ are pulled back and
$\triangleq $ denotes equality holds only on $\Delta$.}
\begin{equation}\label{IHbc}
o^{A}\nabla_{\ub{a}}o_{A}\=0
\hspace{0.2cm}\mbox{and}~~~\iota^{A}\nabla_{\ub{a}}\iota_{A}\=\mu\bm_{a} 
\end{equation}
where, $\mu$ is a real, nowhere-vanishing, spherically symmetric function, and
$\nabla_{a}$ denotes
the unique torsion-free connection compatible with $\sigma_{a}{}^{AA^{'}}$.
\item All equations of motion hold on $\Delta$ and the forms of the fields are
such that $-T_{a}^{b}\l^a$ is causal and $e\=T_{ab}\l^{a}n^{b}$ is spherically
symmetric.
\end{enumerate}

We will study the consequences of these conditions after we have pointed out the
restrictions from $\mathcal{N}=2$ pure Supergravity.

\section{Conditions from $\mathcal{N}=2$ pure supergravity}
The purpose of this section is to establish the compatibility of
the isolated horizon boundary conditions to the conditions
obtained from the KSEs of $\mathcal{N}=2$ supergravity. In other words,
we intend to show that on the horizon, the KSEs can be put in a form which are
precisely the same as the IH boundary conditions\footnote{Since the IH
boundary conditions deal only with the horizon, regardless of structure of the
exterior spacetime, it is enough to check the KSEs on the horizon. However,
in this paper, we intend study the global and completely supersymmetric 
configurations like the Reissner-Nordstrom spacetime and hence we shall also
look for solutions of KSEs in the exterior.}.
This approach was also addressed in \cite{bl1,bl2,l2}. However, for our purpose,
which includes construction of the symplectic structure, deriving the first law
of black hole mechanics and to understand the origin of entropy of the black
holes arising in $\mathcal{N}=2$ supergravity, further details about spacetime
connection and its curvature will be required. This needs
a study of all the constraints available from the KSEs.

Before diving into formal calculations, let us try to 
comprehend the method. As mentioned previously, we are
interested in BPS configurations. In other words, we look for classical
configurations which have some residual supersymmetry and hence satisfy the KSEs
for $\mathcal{N}=2$ supergravity. All such configurations 
(not necessarily solutions of equation of motion) for $\mathcal{N}=2$ have
already been determined and classified in \cite{GH1,Tod1,Tod2} by explicitly
solving the KSEs. It then remains to identify, in the above classification,
weather there exists BPS configuration(s) (\emph{i.e.} member(s) of solution of
KSEs) which also satisfy the isolated horizon boundary conditions. In an 
elaborate way, we can pick each configuration and check weather it actually
satisfies the IH boundary conditions. Alternatively and equivalently, we might
show that (in case when the Killing vector is null, see the paragraph just above
eqn. (\ref{Keqn}) below), the conditions arising out of the KSEs can be put in a
form which will be \textit{exactly} identical to the isolated horizon boundary
conditions. This will easily establish that under these circumstances, there
exists classical configurations of $\mathcal{N}=2$ pure supergravity which will
satisfy the isolated horizon boundary conditions. All such
configurations might not be solutions of equation of motion, only a few will
be. We shall argue that there also exists solutions of equations of
motion, for the theory under consideration, in this space of
configurations. As an example, shall explicitly show that the Reissner-Nordstrom
black hole, which solution of $\mathcal{N}=2$ pure supergravity equations of
motion, solves the KSEs and its horizon satisfies the isolated horizon boundary
conditions. This will be done in two steps: first, we shall show that when there
is a timelike static Killing vector, the configuration obtained by solving the
KSEs is identical to the external region (of the horizon) of the
Reissner-Nordstrom spacetime. Secondly, when the Killing vector is null, the
KSEs can be put in a form identical to the IH boundary conditions
which are also the ones satisfied by the horizon of Reissner-
Nordstrom spacetime. Thus, the Reissner- Nordstrom spacetime will become
consistent with IH fromalism. This will also demonstrate that other black hole
solutions in supergravity theories can be understood using the isolated horizon
formulation.

Let us now look into the KSEs of the $\mathcal{N}=2$ pure supergravity.
This theory has graviton and Maxwell field as the bosonic fields and
two gravitini as their fermionic counterparts \cite{fvn}. We are
interested in the bosonic sector, with the gravitini fields set to zero since
this sector will give us the classical spacetime solutions. Supersymmetry
transformations are generated by gauge-spinor fields
$\epsilon^{\mathbf{A}}_{A}=(\alpha_{A}, \beta_{A})$ where $\mathbf{A}$ is the
internal $O(2)$ index and $A$ is the spinor index. We start with the standard
Killing spinor equations \cite{Tod1,Tod2, bl1, bl2}. 
\begin{eqnarray}\label{kse_Tod}
\nabla_{AA^{'}}\alpha_{B}=\!-\sqrt{2}\:\phi_{AB}\beta_{A^{'}}\nonumber \\
\nabla_{AA^{'}}\beta_{B^{'}}=\!\sqrt{2}\:\bar\phi_{A^{'}B^{'}}\alpha_{A},
\end{eqnarray}
where $\phi_{AB}$ is the anti self-dual part of $F_{ab}$, \emph{i.e.},
\begin{equation}\label{expFab_dyad}
F_{ab}=\sigma_{a}^{AA^{'}}\sigma_{b}^{BB^{'}}F_{AA^{'}BB^{'}}=(\phi_{AB}\,
\epsilon_{A^{'}B^{'}}+
\epsilon_{AB}\,\bar{\phi}_{A^{'}B^{'}}).
\end{equation}

For further calculations, it is useful to define the
function $V=\alpha_{A}\bar\beta^{A}$. Also define the vector
fields \footnote{From now on, we shall omit the soldering form
$\sigma_{a}^{AA^{'}}$. Double spinor indices of the same type will
indicate one spacetime index, for e.g. $Q_{a}\equiv Q_{AA^{'}}$.}:
\begin{equation}\label{LNMdef}
L_{a}\equiv L_{AA^{'}}=\alpha_{A}\bar{\alpha}_{A'},~~~N_{a}\equiv
N_{AA^{'}}=\bar{\beta}_{A}\beta_{A'}~~~~\mbox{and}~~~
M_{a}\equiv M_{AA^{'}}={\alpha}_{A}\beta_{A'}
\end{equation}
We first concentrate on the case $V\ne 0$. It follows from the equations in
(\ref{kse_Tod}) that $J_{a}=(L_{a}-N_{a})$ and $M_{a}$ are local gradients. A
combination of the two vector fields $M_{a}$ and $\bar{M}_{a}$
can be used to define the $(\theta,\phi)$ plane \cite{Tod1,Tod2}. From
(\ref{kse_Tod}), we also get that the vector field $K^{a}=(L^{a}+N^{a})$ is
timelike and Killing:
\begin{equation}\label{K_Killing}
\nabla_{a} K_{b}+\nabla_{b} K_{a}=0
\end{equation}
All the static configurations admitting solution
to the eqn. (\ref{kse_Tod}) are in the Majumdar-Papapetrou class \cite{GH1,Tod1,
Tod2}. If $r_{H}$ defines the horizon, the space of solutions of eqn.
(\ref{kse_Tod}) include the $r>r_{H}$ part of the extremal
Reissner-Nordstrom spacetime. This is the only static solution (in the
Majumdar-Papapetrou class) with a single horizon. It follows that when the
Killing vector is timelike, a solution of the Killing spinor equation is
indeed the spacetime exterior to the Reissner-Nordstrom horizon.

The degenerate sector, $V=0$ is somewhat subtle and needs care.
Before going into that, let us study the $V\ne 0$ case in greater detail.
From equation (\ref{K_Killing}), we get:
\begin{equation}\label{Killing_res}
K^{a}K^{b}\nabla_{b}K_{a}=0
\end{equation}
Since the vector field $J_{a}$ is orthogonal to $K^{a}$ \emph{i.e.}
$K^{a}J_{a}=0$, we get from (\ref{Killing_res}) that:
\begin{equation}\label{Killing_ortho}
K^{b}\nabla_{b}K_{a}= \chi J_{a},
\end{equation}
where $\chi$ is some function\footnote{More generally, since  $M_{a}$
and $\bar{M}_{a}$ are also orthogonal to $K^{a}$, $K^{a}M_{a}=0$ etc. the
equation \ref{Killing_ortho} should be $K^{b}\nabla_{b}K_{a}= \chi J_{a} +
\bar{s}M_{a} +s \bar{M}_{a}$. However, we are working with static solutions
having timelike Killing vector field $K^{a}$. This implies that
$K^{a}\nabla_{a}K_{b}$ will not include the $(\theta,\phi)$ components. In other
words, we can concentrate only on the deformations of the $``(r-t){}"$ plane,
keeping aside the `sphere' ($M_{a}, \bar{M}_{a}$) part}. For $V\ne 0$, the
vector fields $K^{a}$ and  $J^{a}$ can be used to define two orthogonal
directions. Indeed the these can describe the $``(r-t){}"$ plane. A combination
of the other two vector fields $M_{a}$ and $\bar{M}_{a}$ define the
$(\theta,\phi)$ plane.

On the horizon, $K^{a}$  become null and hence, is automatically geodetic:
\begin{equation}\label{K_geod}
K^{b}\nabla_{b}K_{a}= \bar{\chi} K_{a},
\end{equation}
where $\bar{\chi}$ is some function on the horizon. Eqn.(\ref{K_geod}) and
(\ref{Killing_ortho}) together imply that on the horizon, $J^{a}\rightarrow
K^{a}$. Thus the entire $``(r-t)"$ plane degenerates (to a line) on the horizon.
We will then identify the horizon to be the surface where, $J^{a}=K^{a}$ (modulo
rescaling by functions). We shall take this as our criterion for 
defining the horizon $\Delta$. We will see below that in this precise sense,
$V=0$ defines a horizon.

When $V$ is vanishing, $\bar\beta^{A}\=K\alpha^{A}$, where the 
function $K$ on $\Delta$ is such that (see \ref{kse_Tod})
\begin{equation}\label{Keqn}
\alpha_{B}\nabla_{\ub{AA^{'}}}K \=\!\sqrt{2}\:(1+K\bar
K)\bar\alpha_{\ub{A^{'}}}\phi_{\ub{A}B}
\end{equation}
From equation (\ref{LNMdef}), we also obtain that $V=0$ implies $J^{a}=K^{a}$
(modulo rescaling by functions). The configurations which solve the 
Killing spinor equations (\ref{kse_Tod}) for $V=0$ actually describe
null surfaces. The standard coordinate system used in the exterior collapses on
such surfaces. For example, in the Reissner-Nordstrom solution, the $``(r-t)"$
plane degenerates on the horizon. The trick to lift such degeneracy of
coordinate systems is to introduce by hand an auxiliary vector field
\footnote{If $\ell^a$ generates the null surface, one can introduce the
auxiliary null vector field $n^a$ such that $\l.n=-1$.}. We shall use this
option. We introduce two normalised spinors $o_{A}$ and $\iota_{A}$ as
the  such that $\iota^{A}o_{A}\=1$. The spinor $o_{A}$ 
is such that  $o_{A}=e^{ig}\alpha_{A}$, where, $g$ is a real
function on $\Delta$. These spinors can be used to construct a null tetrad basis
$(\l, n,m, \bar{m})$ (compare with eqn. (\ref{lnmdef})).

For further calculations, we need the form of the Maxwell field ($\phi_{AB}$).
Observe that the equation (\ref{Keqn}) can be rewritten as:
\begin{equation}\label{Keqn_transf}
\nabla_{\ub{AA^{'}}}K \=\!\sqrt{2}\, e^{2ig}\:(1+K\bar
K)\, ~\iota^{B}\,\bar{o}_{\ub{A^{'}}}\phi_{\ub{A}B}
\end{equation}
Since, $K$ is a function on $\Delta$, the right hand side of (\ref{Keqn_transf})
can depend on $n_{a}, m_{a}, \bar{m}_{a}$. Then, $\phi_{AB}$ can be of the
form (see eqn. (\ref{lnmdef})) 
\begin{equation}\label{phiab_exp}
\phi_{AB}\=\phi_{0} ~\iota_{A}\iota_{B}+\phi_{1}~(\iota_{A}o_{B}+o_{A}\iota_{B}
)+\phi_{2}~o_{A}o_{B}
\end{equation}
However, we shall see that $\phi_{0}=0$ and there cannot be any term
proportional to $\iota_{A}\iota_{B}$. This is because, the surface
$\Delta$ is null and the energy-momentum tensor ($T_{ab}$) must be such that
the vector field $-T_{a}{}^{b}\l^{a}$ is causal or null. For the Maxwell fields,
the energy-momentum tensor is given by:
\begin{equation}\label{Tabexp}
T_{ab}=\frac{1}{4\pi}\left[F_{ac}F^{c}{}_{b}-\frac{1}{4}g_{ab}F^{2}\right]
\end{equation}
Using the eqns. (\ref{expFab_dyad}), (\ref{Tabexp}) and form of $\l^a$
in terms of spin-dyads (see eqn. (\ref{lnmdef})), the abovementioned
restriction on $T_{ab}$ implies that $\phi_{0}=\bar{\phi}_{AB}~o^{A}o^{B}\=0$.
For future convenience, we shall call $1/2\pi |\phi_{1}|^2\= e$. The other
component of the Maxwell field $\phi_{2}$ will be kept unrestricted.

Let us now determine the constraints on the null-normals on $\Delta$. Using eqn.
(\ref{kse_Tod}) and (\ref{phiab_exp}), we get 
\begin{equation}\label{oeqn}
\nabla_{\ub{a}}~o_{B}\= (\sqrt{2}i\bar{K} \phi_{1}\bar m_{a} -
i\nabla_{\ub{a}}~g)~o_{B}\=:-\tilde{\alpha}_{a}~o_{B}
\end{equation}
From the normalization condition $\iota^{A}o_{A}\=1$ and eqn. (\ref{oeqn}), we
can obtain the action of the gradiant operator on $\iota_{A}$. We restrict the
form to be
\begin{equation}\label{ieqn}
\nabla_{\ub{a}}\iota_{B}\= \tilde{\alpha}_{a}\iota_{B} +\mu \bar m_{a} o_{B},
\end{equation}
where $\mu$ is a function on $\Delta$. With the equations (\ref{oeqn}) and
(\ref{ieqn}) in hand, we can proceed to study their
consequences. Note that these equations are precisely of the
same form as the IH boundary conditions.
Moreover, the constraints on the Maxwell field derived from eqn.
(\ref{phiab_exp}) are such that they satisfy the conditions matter field must
satisfy on an IH (see section \ref{ihbc}).

\section{Consequences of boundary conditions}
In this section, we shall study the kinematical consequences
of the boundary conditions. In what follows, we shall always restrict to
horizons which are spherical, \emph{i.e.} the null surface $\Delta$ will
be foliated by spheres. Using eqn. (\ref{lnmdef}), we find that the equations
(\ref{oeqn}) and (\ref{ieqn}) imply 
\begin{eqnarray}\label{gradeqns}
\nabla_{\ub{a}}\ell^b&\=&\omega^{(\l)}_{a}\l^{b} \label{omega_eqn}\\
\nabla_{\ub{a}}n^b&\=&-\omega^{(\l)}_{a} n^{b} + \mu \bm_{a} m^{b} + \bar\mu
m_{a} \bm^{b} \label{U_eqn}\\
\nabla_{\ub{a}}m^b&\=&V_{a}^{(m)}m^b + \mu m_{a}\ell^b,\label{V_eqn}
\end{eqnarray}
where, $\omega^{(\l)}_{a}=-2 \mathrm{Re}(\tilde{\alpha}_{a})$ and
$V_{a}^{(m)}=-2i\mathrm{Im}(\tilde{\alpha}_{a})$ are one-forms on $\Delta$. 
The superscripts $\l$ and $m$ indicate that $\omega^{(\l)}_{a}$ and
$V_{a}^{(m)}$ depend on the transformation of these vector fields. Also, note
that $V_{a}^{(m)}$ is purely imaginary.

Several consequences follow from these equations \cite{abf,cg_holst}. Firstly,
since $\l^a$ is null (and generates $\Delta$), it is automatically geodetic
and shear-free. Moreover from eqn. (\ref{omega_eqn}), the expansion 
$\theta_{(\l)}$ vanishes on $\Delta$. All these restrictions, the Raychaudhuri
equation for $\l^a$ and the the energy condition further imply that  $\l^a$ is
also shear-free. From eqn. (\ref{omega_eqn}), we get that the null-vector field
$\ell^a$ is a Killing vector \textit{on} $\Delta$,
\begin{equation}
\lie_{\ell}\,g_{\ub{ab}}\=0.
\end{equation}
\textit{i.e.} the IH boundary conditions imply that it is enough to have a
Killing vector only \textit{on} $\Delta$. Further, the volume form of the
$2$-spheres foliating $\Delta$ given by ${}^2\epsilon=im\wedge\bar m$, is also
lie dragged: $\lie_{\ell}\,{}^2\epsilon_{\ub{ab}}\=0$. To see this, use the
Cartan formula $\lie_{\l}\,{}^2\epsilon=d(\l.\,{}^2\epsilon)+
\l.\,d\,{}^2\epsilon$, and the equation (\ref{V_eqn}). The surface gravity of
$\ell^a$ is denoted by $\kappa_{(\ell)}$:
\begin{equation}\label{l_surface_grav}
\l^b\nabla_{b}\l^{a}\=\kappa_{\l}\l^a,
\end{equation}
The equations (\ref{omega_eqn}) and (\ref{l_surface_grav})
together imply that $\kappa_{(\ell)}\=\omega^{(\ell)}_{a}\ell^a$.

The properties of the vector field $n^a$ follow similarly. It is 
twist-free, shear-free, has spherically symmetric expansion $\theta_{(n)}\=2\mu$
and vanishing $\pi (=\ell^{a}\bar m^{b}\nabla_{a}n_{b})$ on $\Delta$. This now
shows that the function $\mu$ is actually related to the expansion of the
vector field $n^a$.

Before proceeding further, let us discuss the issues related to the available
gauge freedom for the spin-dyads on $\Delta$. The most general transformation
that preserves the normalization of the dyad $(\iota_{A}, o_{A})$ is \cite{abf}
\begin{equation}
(\iota^{A}, o^{A})\rightarrow (e^{\Theta-i\theta}\iota^{A}, e^{-\Theta +i\theta}
o^{A}),
\end{equation}
where $\Theta$ and $\theta$ are real functions on $\Delta$. Under
(\ref{lnmdef}), transformations,
the null vectors are
\begin{equation}
\ell^a \rightarrow \xi\ell^a~~~~~~n^{a}\rightarrow
\frac{1}{\xi}n^a
~~~~m^a\rightarrow e^{if}m^a ~~~~\bm^a\rightarrow e^{-if}\bm^a ,
\end{equation}
where, $\xi=e^{-2\Theta}$ and $f=2\theta$ are functions on $\Delta$. The
one-forms $\omega^{(\l)}_{a}$ and $V_{a}^{(m)}$ transform like gauge fields
under rescaling of $\ell^a$ and $m^a$ respectively:
\begin{eqnarray}
\omega^{(\xi\l)}_{a}\=\omega^{(\l)}_{a}+\nabla_{a}ln\,\xi \\
V_{a}^{(fm)}\=V_{a}^{(m)}+i\nabla_{a}\, f
\end{eqnarray}
Consequently the surface gravity also depends on the rescaling of $\l^a$,
$\kappa_{(\xi\ell)}\=\kappa_{(\ell)}+\lie_{\l}\xi$. In this paper, we are 
interested in asymptotically flat global solutions, \textit{i.e} where the
observer one has access to the infinity. In these special cases, the
vector field can always be normalized with respect to infinity. In other 
words, we can set $\Theta=0$ so that there is no scaling ambiguity in the
evaluation of the surface gravity $\kappa_{(\ell)}$. The function $\theta$
however remains unrestricted.

For further calculations, we shall need the curvature of the one-form fields
$\omega^{(\l)}$ and $V^{(m)}$. They are given by \cite{ack,cg_holst}:
\begin{eqnarray}
d\omega^{(\l)}&\=&2\mathrm{Im}(\Psi_{2}){}^{2}\epsilon,\\
dV^{(m)}&\=&\frac{2}{i}(\mathrm{Re}\Psi_{2}-\Phi_{11}
-\frac{R}{24}){}^{2}\epsilon\label{dveqn}
\end{eqnarray}
where, $\Psi_{2}$ and $\Phi_{11}$ are components of the Weyl and the Ricci
tensor respectively (see \cite{Chandra, PR_book}).
The spherical symmetry of the horizon implies that $d\omega^{(\ell)}\=0$
or in other words, $\omega^{(\ell)}$ is a pure gauge and can be made to vanish
by a choice of gauge. We are interested in extremal black holes. These have
vanishing surface gravity. For future calculations, we shall always set
$\omega^{(\ell)}\=0$. Eqn. (\ref{oeqn}) shows that this can be done for some 
special choice of the function $g$.

\section{Laws of black hole mechanics and entropy}
In this section, we shall derive the zeroth and the first law of black hole
mechanics. We will construct the symplectic structure for 
$\mathcal{N}=2$ supergravity with appropriate boundary conditions and 
derive the first law. Thereafter, we will restrict to fixed area part of
the phase space and derive the effective theory residing on the horizon.
This will give us clues to calculate the entropy.

\subsection{The Zeroth Law} We call $\Delta$ a Weakly Isolated Horizon (WyIH) if
$\lie_{\l}\omega^{(\l)}\=0$. The requirement can be justified as follows. The
quantity $\omega^{(\l)}$ is analogue of the extrinsic curvature on the null
hypersurface \cite{afk}. Since $\ell^a$ is a Killing vector on
$\Delta$, the above condition implies that the entire data on the phase space
is lie dragged by $\ell^a$. Using the Cartan equation for lie derivative, it
follows that the surface gravity is constant 
\begin{equation}
d\,\kappa_{(\l)}\=0
\end{equation}
throughout the horizon. For extremal and spherical black holes,
$\omega^{(\l)}\=0$. The restriction of WyIH implies that
the surface gravity $\kappa_{(\l)}$ is zero and remains constant on $\Delta$.

\subsection{The First Law}

For the first law, we need to study the dynamics. For this, we need an action 
which will specify the dynamics and interactions of the geometric and matter
degrees of freedom. We will use the Holst action modified for $\mathcal{N}=2$
supergravity theory \cite{k}.

\subsubsection{Holst Type Modification of $\mathcal{N}=2$ Supergravity Action}
The $\N =2$ supergravity has an Abelian gauge field $\bA$, the tetrad fields
$e_{a}^{I}$ and two superpartner gravitinos. These gravitinos have chiral
projections. One of them, denoted by $\psi_{a}^{\alpha}$, has positive
chirality and the other, denoted by $\psi_{a \alpha}$, has negative chirality.
For writing thr first order action, we also introduce $SO(3,1)$ Lie
algebra-valued connection one form $A_{IJ}$. The modified action for
$\N =2$ supergravity is given by \cite{k}:
\begin{equation}\label{Sholst_action}
S_{SG2}= S_{SSG2}+S_{MSG2}
\end{equation}
where, the term $S_{SSG2}$ is the standard supergravity
action for $\N =2$:
\begin{eqnarray}
S_{SSG2}&=&\int_{\mathcal{M}} d^{4}x e
\left[\frac{1}{2}\Sigma_{ab}{}^{IJ}F_{IJ}{}^{ab}-\frac{1}{4}\bF_{ab}\bF^{ab}
-\frac{1}{2e}\epsilon^{abcd}
(\bar\psi^{\alpha}_{a}\gamma_{b}D_{c}\psi_{\alpha d}-\bar\psi_{\alpha
a}\gamma_{b}D_{c}\psi^{\alpha}_{d})\right ] \nn
&&~~+\left[\frac{1}{2\sqrt
2}\bar\psi^{\alpha}_{a}\bar\psi^{\beta}_{b}\epsilon_{\alpha\beta}(\bar
\bF^{+ab} + \bF^{+ab})
+\frac{1}{2\sqrt 2}\bar\psi_{\alpha a}\bar\psi_{\beta
b}\epsilon^{\alpha\beta}(\bar \bF^{-ab} + \bF^{-ab})\right ]
\end{eqnarray}
where, $\bF=d\bA$, $\bar{\Sigma}_{ab}{}^{IJ}=(e_{a}{}^{I}\wedge e_{b}{}^{J})$
and $F_{IJ}$ is curvature of $A_{IJ}$, \textit{i.e} $F_{IJ}=dA_{IJ}+A_{IK}\wedge
A^{K}{}_{J}$. The
supercovariant field strength is given by 
\begin{equation}
\bar \bF_{ab}=\partial_{[a}A_{b]}-\frac{1}{\sqrt
2}(\bar\psi^{\alpha}_{a}\bar\psi^{\beta}_{b}\epsilon_{\alpha\beta} +
\bar\psi_{\alpha a}\bar\psi_{\beta b}\epsilon^{\alpha\beta})
\end{equation}
and the $+$ and $-$ signs denote the self dual and anti self dual fields,
$\bF_{ab}^{+}=\frac{1}{2}(\bF_{ab}+ {}^{\ast}\bF_{ab})$
and ${}^{\ast}\bF_{ab}=\frac{1}{2e}\epsilon_{abcd}\bF^{cd}$.
The other part of the action  $S_{MSG2}$ is given by:
\begin{eqnarray}
S_{MSG2}&=&\int_{\mathcal{M}} d^{4}x \frac{e}{\gamma}
\left[\frac{1}{2}\Sigma_{ab}{}^{IJ}\bar
F_{IJ}{}^{ab}-\frac{1}{4}\bF_{ab}\bF^{ab} -\frac{1}{2e}\epsilon^{abcd}
(\bar\psi^{\alpha}_{a}\gamma_{b}D_{c}\psi_{\alpha d}+\bar\psi_{\alpha
a}\gamma_{b}D_{c}\psi^{\alpha}_{d})\right] \nn
&&~~-\left[\frac{1}{4e}\epsilon^{abcd}\bar\psi^{\alpha}_{a}\bar\psi^{\beta}_{b}
\bar\psi_{\alpha c}\bar\psi_{\beta d}\right],
\end{eqnarray}
where $\gamma$ is the Barbero-Immirzi parameter. Equations of motion and other
ramifications can be found in \cite{k}.

We are interested in black hole solutions. Moreover, we are restricting to
Reissner-Nordstrom type of configurations. These are global supersymmetric
solutions. From the point of view of classical solutions, the degrees of
freedom of this theory are equivalent to the Holst action
of the Einstein-Maxwell system, which is a consistent truncation of the 
modified Holst action for $\N =2$ supergravity given in
eqn. (\ref{Sholst_action}). In other words, for studying global classical
solutions, we can consistently put the fermion fields to be zero. The KSEs for
(\ref{Sholst_action}) are identical to that of the Einstein-Maxwell system.
In what follows, we shall use the following action:
\begin{equation} \label{EM_theory_action}
S_{H}= \dfrac{1}{16\pi G}\int_{\mathcal{M}}\Sigma_{IJ}\wedge F^{IJ}-\dfrac{1}{16
\pi G \gamma}\int_{\mathcal{M}}e^I \wedge e^J\wedge
F_{IJ}-\frac{1}{8\pi}\int_{\mathcal{M}}{\bf F}\wedge{}^{*}{\bf F}
\end{equation}
where, $\Sigma_{IJ}=\frac{1}{2}\epsilon_{IJKL}e^K \wedge e^L$ is
a $2$-form and $\epsilon_{IJKL}$ is the completely antisymmetric tensor in
internal space. Variation of the action with respect to the connection $A_{IJ}$
leads to:
\begin{equation}\label{dsigma1} D\Sigma_{IJ}=0
\end{equation}
It then follows from (\ref{dsigma1}) that $A_{IJ}$ is
the spin connection. Then the variation of the action (\ref{EM_theory_action})
with respect to the tetrads $e_{I}^{a}$ give Einstein-Maxwell equations
\cite{al}. The boundary terms that arise from the variation of the action get
contributions from the inner and outer boundaries. However, IH (or SWIH)
boundary conditions  and asymptotic flatness ensure that these boundary terms
vanish, making the action principle well-defined \cite{cg_holst}.

We will need to  specify the form of the Lorentz Lie-algebra valued connection
one-form $A_{\ub{a}{IJ}}$ on $\Delta$. Introduce a fixed
set of internal null vectors $(\l^{I}, n^{I}, m^{I}, \bar m^{I})$ on $\Delta$
such that $\ell^{I}n_{I}=-1=-m^{I}\bm_{I}$ while the
other inner products vanish. These internal vectors are such that
$\partial_{a}(\l^{I}, n^{I}, m^{I}, \bar m^{I})=0$.
Given these internal null vectors and the tetrad $e^I{}_{a}$, we
can construct the null vectors $(\l^{a}, n^{a}, m^{a}, \bar m^{a})$ through
$\l_a =e^I{}_{a}\l_{I} $. We can use these information to find the connection. 
To do this, we first note that since the internal null vectors
are fixed, ($\partial_{a}(\l^{I}, n^{I}, m^{I}, \bar m^{I})=0$), for the
internal vector $\ell^{I}$ we get 
\begin{equation}\label{eq-conn-cal}
\nabla_{\ub{a}}\l_{I}\= A_{\ub{a}I}{}^{J}\l_{J}.
\end{equation}
Similar expressions can be obtained for the other internal vectors. Using the
equations (\ref{gradeqns}), the full connection turns out to be:
\begin{equation} \ub{A}_{IJ}\= 2\bar\mu m~\l_{[I}\bm_{J]}+ 2~\mu \bm~\l_{[I}
m_{J]}+ 2~V^{(m)}~m_{[I}\bm_{J]}
\end{equation}
We define the following Lie algebra valued connection $A^{(H)}_{IJ}$ for ease of
further computation \cite{al,cg_holst} 
\begin{eqnarray} 
\ub{A}^{(H)}_{IJ}&=&\frac{1}{2}\left( A_{IJ}-\frac{\gamma}{2}
\epsilon_{IJ}{}^{KL}A_{KL}\right)\nonumber \\
&\=& ~V^{(m)}~\left(i\gamma \l_{[I}n_{J]} +  m_{[I}\bm_{J]}~\right)
+ \bar\mu m~\l_{[I}\bm_{J]}~\left(1 +i\gamma\right)\nn
&&+ \mu \bm~\l_{[I}m_{J]}~\left(1-i\gamma\right),
\end{eqnarray}
We shall also need the expression of the product of tetrads on the horizon
$\Delta$. It is easily determined to be:
\begin{equation}\label{tetrad_prod_exp}
\ub{e}^{I}\wedge \ub{e}^{J}
=-2~n\wedge m~\l^{[I}\bar m^{J]} -2~n\wedge \bar m~\l^{[I}
m^{J]} +2i~m^{[I}\bar m^{J]}~{}^2\mbf{\epsilon}
\end{equation}

\subsubsection{Symplectic structure}
Given the lagrangian $4$-form, there exists specific prescription for
constructing the symplectic structure on the space of solutions
\cite{w,wi,abr,lw}. One obtains on-shell, the symplectic one-form $\Theta$ (a
spacetime $3$-form) from the variation of the Lagrangian, $\delta L= d\Theta
(\delta)$ where $\delta$ is an arbitrary vector field in the phase space.
For the case in hand, we get 
\begin{equation}
~\Theta(\delta)=-\frac{1}{8\pi G\gamma}~\delta(e^{I}\wedge
e^{J})\wedge A^{(H)}_{IJ}+\frac{1}{4\pi}\delta{\bA}\wedge {}^{*}\bF
\end{equation}
From $\Theta(\delta)$, one then constructs the symplectic current 
$J(\delta_{1},\delta_{2})=\delta_{1}\Theta(\delta_{2})-
\delta_{2}\Theta(\delta_{1})$. This is closed on-shell and integrating over
the entire spacetime, we get (see figure 1):
\begin{eqnarray}\label{Jdeltaeqn}
(\int_{M_{+}}- \int_{M_{-}})J(\delta_{1},\delta_{2})&\= &\frac{1}{8\pi G
\gamma}\int_{\Delta} [\delta_{1}{}^2\mbf{\epsilon} \wedge
\delta_{2}(iV^{(m)})-(1\leftrightarrow 2)]\nn
&+&\frac{1}{4\pi}\int_{\Delta} [\delta_{1}\bA\wedge
\delta_{2}{}^{*}\bF-(1\leftrightarrow 2)]
\end{eqnarray}
To construct the symplectic structure, we must be
careful that no data flows out of the phase space because of our choice of
foliation. To ensure this, we will check that the symplectic structure is
independent of the choice of foliation. We introduce potentials
$\mu_{(m)}$ and $\varphi_{(\l)}$.
\begin{equation}
\lie_{\l}\,\mu_{(m)}\=i\l^a V^{(m)}_{a} ~~~~\mbox{and} ~~~
~\lie_{\l}\,\varphi_{(\l)}=\Phi_{(\l)}\=-\ell^{a}{\bf A}_{a}
\end{equation}
A straightforward calculation shows that (see \cite{cg1,cg_holst}):
\begin{eqnarray}
(\int_{M_{+}}- \int_{M_{-}})J(\delta_{1},\delta_{2})&\= &\frac{-1}{8\pi G
\gamma}(\int_{S_{-}}-
\int_{S_{+}})\{\delta_{1}{}^2\mbf{\epsilon}~\delta_{2}\,\mu_{(m)}-
(1\leftrightarrow 2)\}\nn
&-&\frac{1}{4\pi}(\int_{S_{-}}-
\int_{S_{+}})\!\big[\delta_1\!*{\bf F}\,\delta_2
 \varphi_{(\ell\,)}-(1\leftrightarrow 2)\big].
\end{eqnarray}
So only a special combination of the bulk and boundary symplectic current is
independent of the choice of foliation. The symplectic structure is
that of a Einstein- Maxwell system (see \cite{cg_holst} for detailed
derivation):
\begin{eqnarray}\label{Palatini_1}
\Omega(\delta_{1}, \delta_{2}
)&=&\frac{1}{8\pi G\gamma}\int_{M}\left[ \delta_{1}(e^{I}\wedge
e^{J})~\wedge\delta_{2}A^{(H)}_{IJ} - (1\leftrightarrow 2)\right] -\frac{1}{8\pi
G\gamma}\oint_{S_{\Delta}}\left[
\delta_{1}{}^2\mbf{\epsilon}~\delta_{2}\mu_{(m)}-(1\leftrightarrow 2)
\right]\nn
&&+\frac{1}{4\pi}\!\int_M\!\big[\delta_1\!*{\bf F}\wedge\delta_2{\bf A}
       -(1\leftrightarrow 2)\big]-
 \frac{1}{4\pi}\oint_{S_\Delta}\!\big[\delta_1\!*{\bf F}\,\delta_2
 \varphi_{(\ell\,)}-(1\leftrightarrow 2)\big]\;.
\end{eqnarray}
\begin{figure}[h] \label{f1}
  \begin{center}
  \includegraphics[height=4.0cm]{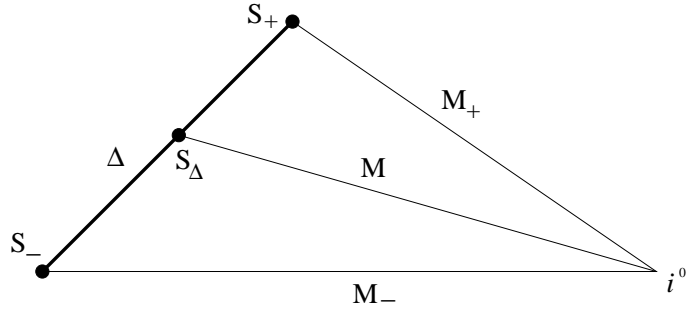}
  \caption{$M_\pm$ are two partial Cauchy surfaces enclosing a  region of
space-time and intersecting $\Delta$ in the $2$-spheres $S_\pm$ respectively
and extend to spatial infinity $i^o$. Another Cauchy slice M is drawn which
intersects $\Delta$ in $S_\Delta$}
  \end{center}
\end{figure}

The first law can now be derived using this symplectic structure
(\ref{Palatini_1}). Let us understand the conceptual basis of this proof.
IH is a local definition of a horizon and the first law is expected to relate
variations of local quantities that are defined only at the horizon without any
reference to the rest of the spacetime \footnote{For Reissner- Nordstrom like
global configurations, the first law will be valid for the entire spacetime.}.
For example, the surface gravity $\kappa_{(\xi\l)}$ is defined locally at the
horizon. For the first law, the IH formalism enables us to define local energy
(for horizons carrying other charges, such as angular momentum, electric
potential etc., we must also provide local definitions for them). In spacetime, 
energy is associated with a timelike Killing vector field. Given any vector
field $W$ in spacetime, it naturally induces a vector field $\delta_{W}$ in the
phase space. The phase space vector field $\delta_{W}$ is the generator of time
translation in the phase space. If time translation is a canonical
transformation in the phase space then $\delta_{W}$ defines a Hamiltonian
function $H_{W}$. So to find out the Hamiltonian function associated with
energy, we must look for phase space transformations that keep the symplectic
structure invariant (canonical transformations). The vector fields tangent to
these canonical flows are the Hamiltonian vector fields. To check wheather a
vector field $\delta_{W}$ in the phase space is Hamiltonian, one constructs a
one-form $X_W$ where $X_{W}(\delta):=\Omega(\delta,\delta_{W})$, where
$\delta_{W}$ is the lie flow $\lie_{W}$ generated by the spacetime vector field
$W^a$ when tensor fields are varied. The necessary and sufficient
condition for the vector field $\delta_{W}$ to be a \emph{globally Hamiltonian}
vector field is that the one-form $X_{W}$ is to be \emph{exact},
$X_{W}=\mathbf{d}H_{W}$ where $\mathbf{d}$ is the exterior derivative in phase
space and $H_W$ is the corresponding Hamiltonian function. In other words, the
vector field $\delta_{W}$ is globally Hamiltonian if and only if
$X_{W}(\delta)=\delta H_{W}$ for any vector field $\delta$
in the phase space. The vector fields $W^a$ are also restricted by the condition
that it should be tangential on $\Delta$. Now being a null surface, the WIH has
only three tangential directions, one null and the two other spacelike. The
closest analog of `time' translation on WIH is therefore translation along the
null direction. It is generated by the vector field $[\l^a]$. For global
solutions this null normal vector field becomes timelike outside the horizon and
is expected to match with the asymptotic time-translation for asymptotically
flat spacetimes.

Using the above considerations, the first law of supersymmetric horizons turns
out to be:  
\begin{equation}\label{1stlaw}
X_{\l}{(\delta)}\= \Phi_{(\ell\,)}\delta Q_\Delta\; +\delta E_{(\l)},
\end{equation} 
where $E_{(\l)}$ is the ADM energy obtained when $\l^a$ matches with the time
translation at infinity and $Q\=-(1/4\pi)\oint_{S_{\Delta}}{}^{*}F$ is the
charge of the electromagnetic field on the horizon.. 
The right hand side of (\ref{1stlaw}) is an exact variation if and only if
$\Phi_{(\ell\,)}$ is a function of $Q_\Delta$ alone.
The phase space is characterized by charge and so $\Phi_{(\ell\,)}$ is a
function of $Q_\Delta$. Define a quantity $E_\Delta$ where
\begin{equation}\label{1stlaw}
\delta E_{\Delta}\= \Phi_{(\ell\,)}\delta Q_\Delta\;
\end{equation} 
such that $H_{\l}=E_{(\l)}-E_{\Delta}$ where $H_{\l}$ is the associated
Hamiltonian function $X_{\l}(\delta)=\delta H_{\l}$. It is natural to
interprete $E_\Delta$ as the locally defined energy of the WIH and
(\ref{1stlaw}) as the first law of the WIH. The quantity $H_{\l}$ 
receives contributions both from the bulk as well as the boundary symplectic
structures and stands for the energy of the region between the WIH and the
spatial infinity. The ADM energy $E_{(\l)}$ is the sum total of these two
energies. For the global solutions we are interested in, it is well known that
the first law is equivalent to: 
\begin{equation}\label{1stlaw}
\delta E_{ADM}\= \Phi_{(\ell\,)}\delta Q_\Delta\;
\end{equation} 
\textit{i.e.}, $E_{\Delta}=E_{\mathrm{ADM}}$. To see this, observe that
for all global solutions, $H_{\ell}=0$. This is because when there is a global
Killing vector field, $\delta_{\l}$ induces infinitesimal gauge transform and is
thus a gauge direction,
\begin{equation}
\Omega(\delta, \delta_{\l})=\delta H_{\l}=0,
\end{equation}
for all $\delta$ on the phase space. So, for any connected component of the
phase space consisting of the spacetimes with global Killing vector field,
$H_{\ell}$ is a constant. This constant can only be some spacetime quantity and
can be the cosmological constant. For the present case, the cosmological
constant vanishes and hence $H_{\ell}$ vanishes too and hence the
energy measured at $\Delta$ is same as that measured by any ADM observer.
This means that this first law is exactly equivalent
to that for the event horizons, with ADM replaced by $\Delta$ in (\ref{1stlaw}).
This is a consistency check for the IH formulation. 

\subsection{Chern-Simons theory and entropy}

In the introduction, we have said that the effective theory residing on the
horizon can only be a topological theory. In this section, we shall outline the
derivation of the Chern-Simons theory on $\Delta$. Detail calculations are
similar to the ones in \cite{cg_holst}.

Let us now restrict to fixed area and fixed charge phase space. Define the
connection component $V^{(H)}=iV^{(m)}/2$. In this case of spherical symmetry,
it can be shown that the Gauss-Bonnet theorem implies that the equation
(\ref{dveqn}) reduce to \cite{ack, cg_holst}
\begin{equation}\label{dvheqn}
dV^{(H)}\=-{}^2\epsilon~[2\pi/\mathcal{A}^{s}_{\Delta}]
\end{equation}
This condition is also called the \emph{quantum horizon condition}. The
subscript $s$ indicates that we are in spherically symmetric phase space.
Putting equation (\ref{dvheqn}) in (\ref{Jdeltaeqn}) and integrating by
parts, we see that:
\begin{eqnarray}
\Omega(\delta_{1}, \delta_{2})&=&\frac{1}{16\pi G\gamma}\int_{M}\left[
\delta_{1}(e^{I}\wedge
e^{J})~\wedge\delta_{2}A^{(H)}_{IJ} -\delta_{2}(e^{I}\wedge
e^{J})~\wedge\delta_{1}A^{(H)}_{IJ} \right] \nn +&{}&\frac{1}{8\pi G
\gamma}\frac{\mathcal{A}^{s}_{\Delta}}{\pi}\int_{S}\{
\delta_{1}V^{(H)}{}^{g}\wedge \delta_{2}V^{(H)}{}^{g}\}
-\frac{1}{4\pi}\!\int_M\!\big[\delta_1\!*{\bf F}\wedge\delta_2{\bf A}
       -(1\leftrightarrow 2)\big],
\end{eqnarray}
where $V^{(H)}{}^g=V^{(H)}+d\mu_{(m)}/2$. Note that the
Maxwell field does not give any contribution to the entropy (see
\cite{cg_holst}). The boundary symplectic structure turns out to be that of
$U(1)$ Chern-Simons theory. The level of the theory $k=\mathcal{A}^{s}_{\Delta}/
4\pi G \gamma$ takes integer values on quantization.

The entropy of the horizon can be obtained by quantization of $U(1)$
Chern-Simons theory and thence counting states. The details of the quantization
technique and various ramifications have been calculated in details \cite{abck}.
The counting of states and entropy computation was first
done in \cite{abck}. Better state counting methods have since been proposed
\cite{dl,mei,gm1} and the one put forward in \cite{gm2} has carefully
reconsidered some intricacies in the counting. The essential idea
is the following: Consider a horizon of area $\mathcal{A}^{s}_{\Delta}$. To
compute the entropy, those states are relevant which satisfy the quantum horizon
condition and have the fixed area of value $\mathcal{A}^{s}_{\Delta}$. Entropy
is obtained by taking logarithm of this value. The detailed counting of the
microscopic quantum states of black hole is based on loop quantum gravity. It is
proposed that the states are characterized by means of spin network basis
\cite{al}. If an edge with lebel $j_{i}$ ends at the horizon
$S_{\Delta}$, it creates a \emph{puncture} with label $j_{i}$. The area of the
horizon will be given by the value $8\pi \gamma
L^{2}_{P}\sum_{i}\sqrt{j_{i}(j_{i}+1)}$, $L_{P}$ being
the Planck length. The punctures are also lebelled by the half-integers $m_{i}$
where $-j_{i}\le m_{i}\le j_{i}$. The quantum horizon condition relates this
eigenstates to that of Chern-Simons theory. The requirement that the horizon is
a sphere imposes the constraint
$\sum_{i}m_{i}=0$. Thus the quantum state associated with cross-section of
horizon are characterized by punctures and spin quantum numbers $j,m$ associated
to each punctures lebel the states. Counting of states establishes that the
entropy is indeed proportional to the area of the horizon.

\section{Discussions}
The objective of this paper was to introduce a new way of calculating the
entropy of extremal black holes in supergravity theories. We observe that the
standard Wald formulation \cite{w,wi} fails to address the issue of entropy for
extremal black holes. This is because the formulation depends on the existence
of bifurcation spheres which are absent for these special black holes (and also
for black holes formed out of collapse). It is then becomes necessary to
formulate new ways to address this problem. Instead of modifying the Wald's
Killing horizon (KH) formulation, we reconsidered the isolated
horizon (IH) formulation of black hole horizon. We matched the
boundary conditions precisely and showed that it is possible to
include the black holes arising in pure $\mathcal{N}=2$ supergravity in the
space of solutions this theory with IH as an inner boundary. Moreover,
we proved the laws of black hole mechanics for these black holes and then went
on to show that tye entropy of these black holes can be easily determined by
quantizing the effective Chern-Simons theory that resides on the inner
boundary of these black holes.

The advantage of this framework is that it doesnot require the entire spacetime
to be supersymmetric. It is very much a possibility that the spacetime
just outside the horizon is non-supersymmetric (in the sense that there are no
Killing spinors that generate supersymmetry as isometries) because of presence
of time dependent fields like electromagnetic and gravitational while only the
horizon itself is supersymmetric (\emph{i.e.} the horizon supports some Killing
spinors) because the horizon is in equilibrium. So, this formulation admits a
larger class of spacetime in its phase-space than the KH or the event horizon
formulation which require  some or the entire spacetime to be
supersymmetric respectively. The laws of black hole mechanics thus proved on the
larger phase space can encompass solutions which were otherwise difficult to
address. Secondly, the method of determining the entropy is direct. It does not
depend on the near-horizon/asymptotic structures (as is done for example in
Kerr-CFT approach \cite{str}) but is based on the quantization of the horizon
topological theory induced as a result of the bulk-boundary gravitational
interaction.

The present method however can be extended in various directions. Firstly, the
present calculation is restricted to black holes in pure $\mathcal{N}=2$
supergravity and can be repeated for black holes in extended supergravity.
Secondly, black holes in higher dimensions are becoming more and more important.
It will be an interesting problem to address this method for higher dimensions.

\section*{Acknowledgments}
The author thanks A. Ghosh for discussions and encouragement at various stages
of the work. He also thanks P. Majumdar for encouragement. The initial stage of
the work began at IFT, UAM/CSIC Madrid. The author 
thanks T. Ortin and P. Meessen at IFT for stimulating discussions on reference
\cite{Tod1}, black holes and various other issues. He also thanks I. Booth for
comments.
\bibliography{bibtex_jhep}
\bibliographystyle{JHEP}
\end{document}